\documentclass{aa}  

\usepackage{graphicx}
\usepackage{txfonts}
\usepackage[caption=false]{subfig}
\usepackage{ulem}
\usepackage{color}

\bibpunct{(}{)}{;}{a}{}{,} 

\begin{document} 

    \title{Neural network reconstruction of the dense matter equation of state derived from the parameters of neutron stars}
    \titlerunning{NN reconstruction of the EOS from NS parameters}

   \author{F. Morawski\inst{1}
          \and
          M. Bejger\inst{1}
          }

   \institute{Nicolaus Copernicus Astronomical Center, Polish Academy of Sciences, Bartycka 18, 00-716, Warsaw, Poland\\
              \email{fmorawski@camk.edu.pl}
             }

   \date{\today}

    \abstract
    {Neutron stars are currently studied with an rising number of electromagnetic and gravitational-wave observations, which will ultimately allow us to constrain the dense matter equation of state and understand the physical processes at work within these compact objects. Neutron star global parameters, such as the mass and radius, can be used to obtain the equation of state by directly inverting the Tolman-Oppenheimer-Volkoff equations. Here, we investigate an alternative approach to this procedure.}
    {The aim of this work is to study the application of the artificial neural networks guided by the autoencoder architecture as a method for precisely reconstructing the neutron star equation of state, using their observable parameters: masses, radii, and tidal deformabilities. In addition, we study how well the neutron star radius can be reconstructed using only the gravitational-wave observations of tidal deformability, that is, using quantities that are not related in any straightforward way.}
    {The application of an artificial neural network in the equation-of-state reconstruction exploits the non-linear potential of this machine learning model. Since each neuron in the network is basically a non-linear function, it is possible to create a complex mapping between the input sets of observations and the output equation-of-state table. Within the supervised training paradigm, we construct a few hidden-layer deep neural networks on a generated data set, consisting of a realistic equation of state for the neutron star crust connected with a piecewise relativistic polytropes dense core, with its parameters representative of state-of-the art realistic equations of state.}   
    {We demonstrate the performance of our machine-learning implementation with respect to the simulated cases with a varying number of observations and measurement uncertainties. Furthermore, we study the impact of the neutron star mass distributions on the results. Finally, we test the reconstruction of the equation of state trained on parametric polytropic training set using the simulated mass--radius and mass--tidal-deformability sequences based on realistic equations of state. Neural networks trained with a limited data set are capable of generalising the mapping between global parameters and equation-of-state input tables for realistic models.}{}
   
   \maketitle

\section{Introduction}
\label{sec:intro}

Neutron stars (NS) are currently the best astrophysical sites for studying the details of dense matter physics in conditions that are inaccessible for terrestrial experiments (see e.g. \citealt{HaenselPY2007} for a textbook introduction) Specifically, this refers to the equation of state (EOS) of dense, cold, neutron-rich matter at densities many times higher than the nuclear saturation density $\rho_s\simeq 2.7\times 10^{14}\ \mathrm{g/cm^3}$, corresponding to the nuclear saturation baryon density $n_s{\simeq}\,0.16\,\mathrm{fm^{-3}}$. 

Because the complete theory of many-body nuclear interactions is not known in full, recent efforts have been focussed on inferring the EOS from astrophysical observations of NSs. Recent observations include the NICER simultaneous measurements of the mass and radius of PSR J0030+0451 \citep{Riley2019,Miller2019}, the 170817 binary NS inspiral detection and parameter estimation done by the LIGO and Virgo Collaborations \citep{2017PhRvL.119p1101A,PhysRevX.9.011001,PhysRevLett.121.161101}, including the measurement of the masses and tidal deformability of the system components, accompanied by the observations of high-energy photons by the Fermi and INTEGRAL satellites \citep{2017ApJ...848L..13A}, as well as several observations of massive ${\simeq}2 M_\odot$ NSs \citep{Demorest2010,Fonseca2016,Antoniadis2013,Cromantie2020}. These observations provide indirect but nevertheless informative answers to the question  of how  compact objects are structured and, hence, the nature of their internal composition. The procedure is based on solving the equations of stellar structure, typically the Tolman–Oppenheimer–Volkoff (TOV) equations \citep{Tolman1939,OppenheimerV1939} for an assumed EOS (or class of EOSs) to subsequently compare the global observable NS parameters, such as gravitational mass $M$, radius $R,$ and, recently, tidal deformability $\Lambda$ as well \citep{FlanaganH2008,VanOeverenF2017} to observed values; in the simplest case of the TOV equation, there is a strong relation between the sequence of global NS parameters (EOS functionals), such as $M(R)$ or $M(\Lambda)$, and the pressure-density $p(\rho)$ relation defining the EOS. Therefore, given a set of astrophysical measurements, it is possible, in principle, to recover the EOS by inverting the TOV equations. In reality, however, astrophysical observations are affected by measurement errors and they are not distributed optimally in the parameter space, meaning that an observer doesn't have any freedom in selecting the intrinsic parameters, such as the mass, $M$, of the observed object, to optimally cover the range of pressure and density so that the part of the EOS relation that is of interest may be revealed. 

 The most common strategy in the estimation of EOS utilises Bayesian inference, which is based on the inversion of the TOV equations and a limited number of observations. Examples of this approach were recently presented in the following works: \citet{Steiner_2010, Steiner_2013,Raithel_2016,Holt_2019,Fasano_2019,PhysRevD.100.103009} and \citet{Traversi2020BayesianIO}. Here, instead of directly inverting the TOV equations, we study an alternative approach based on a machine-learning (ML) artificial neural network (ANN), inspired by the autoencoder (AE) architecture \citep{10.5555/2987189.2987190,Goodfellow2016}. Similar machine learning techniques applied to results of numerical simulations and measurements currently make up a field of active research; for example, \cite{2019arXiv191101496H} show that the final mass and spin of a Kerr black hole can be predicted from the initial values of parameters of black hole components. Specifically, this has been explored in the field of the NS EOS, \citet{2018PhRvD..98b3019F,PhysRevD.101.054016} presented an application of a feed-forward neural network to infer the EOS from NS mass-radius measurements, whereas \citet{2019arXiv191005554F} compare machine-learning neural networks and support vector machine regression methods in unveiling the nuclear EOS parameters from NS observations. 

Here, we study the application of ML to infer the dense-matter EOS pressure-density $p(\rho)$ relations from a simulated set of NS observations, using a neural network trained on a purposefully simple data set, based on piecewise relativistic polytrope EOS. We performed the analysis using simulated data containing electromagnetic as well as gravitational-waves observables: radii, masses, and tidal deformabilities, applying the current knowledge of the NS mass distribution function, and varying the number of simulated observations and measurement uncertainties. While trained and tested on the piecewise relativistic polytropic EOS data set, our ML model was also validated on realistic EOS examples: it successfully recovers the SLy4 EOS \citep{DouchinH2001} as well as the APR EOS \citep{PhysRevC.58.1804} and the BSK20 EOS \citep{PhysRevC.82.035804}. Additionally, we show the ANN network is flexible enough to generalise the mapping of the mass-radius $M(R)$ relation from the mass--tidal-deformability $M(\Lambda)$ relation, effectively allowing for the possibility of inference of the NS radius from several GW-only measurements. 

The outline of the article is as follows. In Sect.~\ref{sec:ml}, we discuss the choice of the machine learning algorithms used. Section \ref{sec:prep} is devoted to the description of the input and output data generation procedures, with a particular emphasis on the EOS and the NS structure. Section~\ref{sec:results} contains results of the neural-networks estimation of the dense-matter EOS from NS observables: $M(R)$ and $M(\Lambda)$. We discuss the results in Sect. \ref{sec:discussion}. We conclude in Sect.~\ref{sec:summary} with a summary of our study.

\section{Machine learning}
\label{sec:ml}

The machine learning field of computer science is based on the premise that algorithms can learn from examples in order to solve problems and make predictions without needing to be explicitly programmed \citep{Samuel:1959}. Among the many ML algorithms, the ANN  currently belong to the most popular. Complex ANN consisting of many neurons combined with various training algorithms (such as the backpropagation and stochastic gradient descent - for textbook review, see e.g. \citealt{Goodfellow2016} and references therein) are able to capture complicated non-linear relationships in the data by composing hierarchical internal representations. The complex (in other words, deeper) the algorithm is, the more abstract features it can, in principle, learn from the data.

The main motivation for employing ANN in our project is associated with non-linear potential of the this ML model. Since each neuron in the network is basically a non-linear function, it is possible to create a complex mapping between the input and the output of the model. This characteristic is necessary for the estimation of EOS based on observables, even when excluding uncertainties since the analytical relations between the input and output parameters are non-linear.

The input to our model are astrophysical measurements of NS parameters (presented to the ANN as two arrays of $M$, $R$ or $M$, $\Lambda$ concatenated into one), whereas at the output we obtain an array of similar shape (concatenated $p$, $\rho$ values). Further in text, we present additional project in which we reconstruct radius based on gravitational observables ($M$ and $\Lambda$ concatenated into one vector), the ANN output consists only of radius values. By definition, the size of the output is half the size of the input.

\section{Data preparation}
\label{sec:prep}

In this section, we describe the design of parametric EOSs and methods used to obtain the stellar parameters.  

\subsection{Equations of state and stellar structure} 
\label{sec:eos_struct} 

\begin{figure}
    \includegraphics[width=0.45\textwidth]{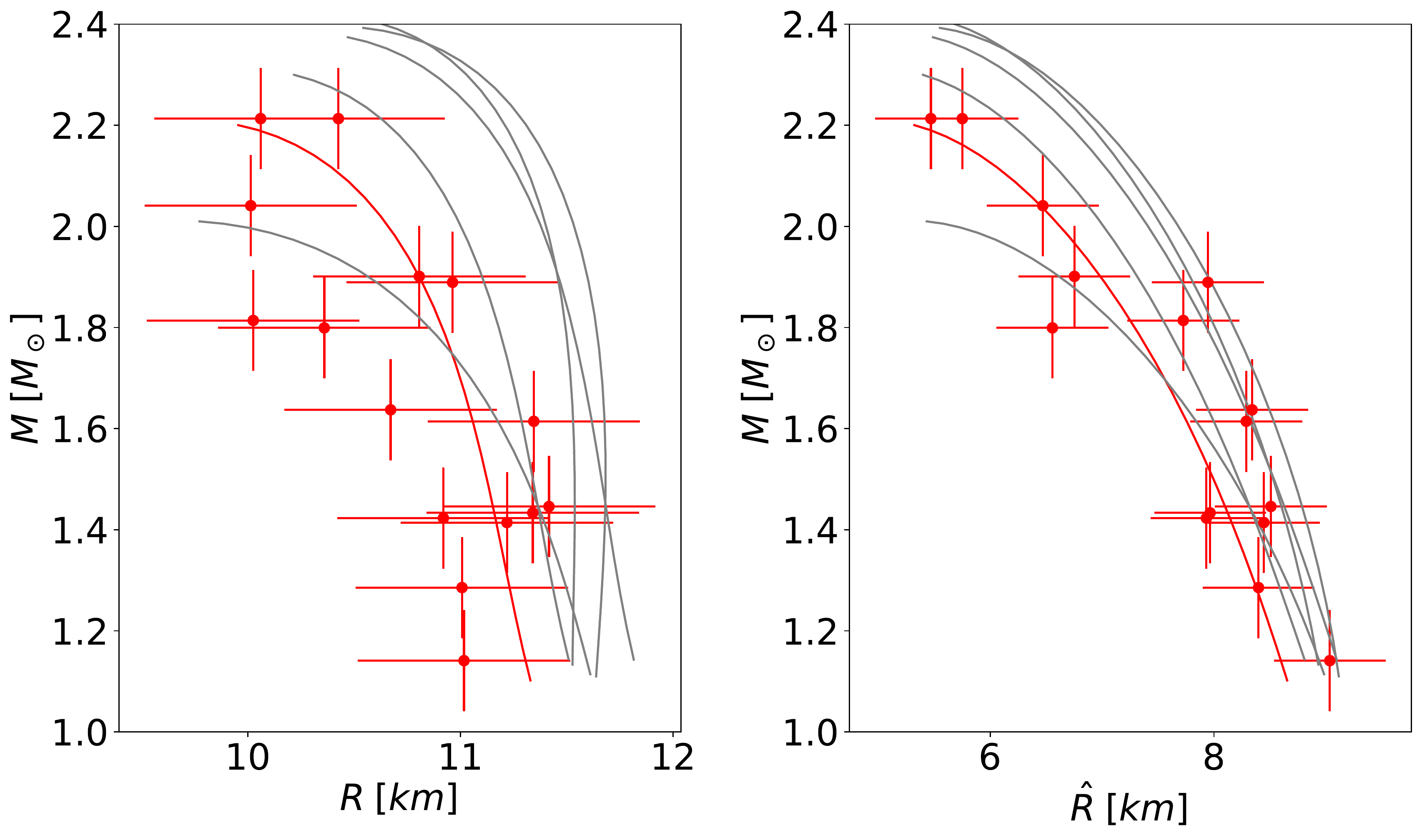}
    \vskip 10pt 
    \hskip 10pt
    \includegraphics[width=0.45\textwidth]{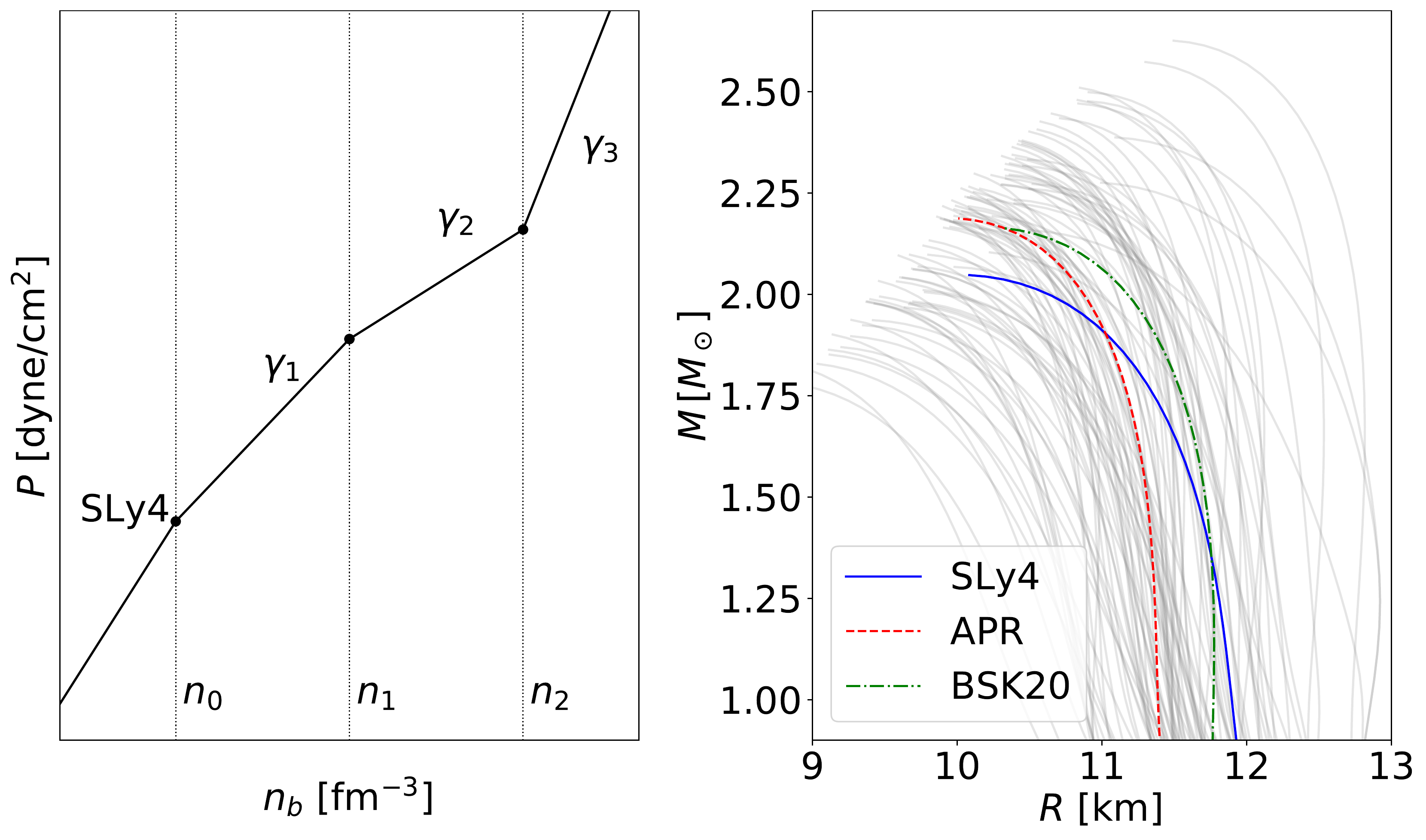}
    \caption{Top panels: left  - Sample $M(R)$ relations; right  - Corresponding $M(\hat{R})$ relations. Red curve and corresponding observations: example of a training datum, consisting on 20 points from the $M(R)$ and $M(\hat{R})$, selected assuming random normal distribution with the mean value equal to true values and standard deviations $\sigma_M = 0.1\ M_\odot$, $\sigma_R = 1.0$ km and $\sigma_{\hat{R}} = 1.0$ km. The configurations with the same $M$ and $R$ have, in general, different $\hat{R}$. 
    Bottom panels: left  - Schematic of a model EOS, composed in the low-density part from the realistic crust of the SLy4 EOS \citep{HaenselP1994,DouchinH2001} and piecewise relativistic polytropes \citep{Tooper1965}, right  - Mass-radius $M(R)$ relations generated using the piecewise relativistic polytrope model (thin solid grey curves) and astrophysical models of NS sequences, based on the following EOSs: the SLy4 EOS (solid blue curve), the APR EOS (dashed red curve) and the BSK20 EOS (dash-dotted green curve).} 
    \label{fig:eos_mr_mrhat}
\end{figure}

In order to cover a sufficiently-broad and representative space of solutions corresponding
to $M(R)$ and $M(\Lambda)$ sequences, we employ the following simplified, parametric approach to the EOS. We assume the measurement of the low-density part
of the EOS and adopt the SLy4 EOS description of \cite{HaenselP1994} and
\cite{DouchinH2001} up to some baryon density $n_0$, comparable to and typically
larger than the nuclear saturation density ($n_s\equiv 0.16$ fm$^{-3}$). At the
$n_0$ a relativistic polytrope \citep{Tooper1965},
\begin{eqnarray} 
  p(n)=\kappa n^{\gamma},\quad \rho c^2 = \frac{P(n)}{\gamma - 1} + nm_{b}c^2, 
  \label{eq:polytrope} 
\end{eqnarray} 
replaces the SLy4 EOS. For each polytrope, the pressure $p$ and the mass-energy density $\rho c^2$ are defined using the pressure coefficient $\kappa$, the polytropic index $\gamma$ responsible for the stiffness of the matter, and the mass of the baryon $m_{b}$. We select the $\gamma$ index as a parameter of choice; consequently, $\kappa$ and $m_{b}$ are determined by demanding the chemical and mechanical equilibrium at $n_0$. The first polytrope with $\gamma_1$ ends at some density, $n_1 > n_0$, where a second relativistic polytrope with $\gamma_2$ is attached, and continues until $n_2$, where a polytrope with $\gamma_3$ starts. The bottom-left panel of Fig.~\ref{fig:eos_mr_mrhat} shows a schematic plot of the EOS. The parameter ranges are collected in Table ~\ref{tab:params}. 

\begin{table}
  \centering 
  \begin{tabular}{| c | c | c | c | c | c | c |} 
    \hline 
      & $n_0$ [fm$^{-3}$] & $\gamma_1$
      & $n_1$ [fm$^{-3}$] & $\gamma_2$
      & $n_2$ [fm$^{-3}$] & $\gamma_3$
      \\ \hline\hline 
    min & 0.1 & 2.5 & $n_0$ & 2.0 & $n_1$ & 3.0 \\ \hline 
    max & 0.2 & 3.5 & 0.3 & 2.5 & 0.4 & 4.0 \\ \hline 
  \end{tabular}
\vskip 1em  
\caption{Ranges of piecewise polytrope EOS parameters used in the study: $n_0$, $n_1$ , and $n_2$ are the baryon densities at which the relativistic polytropes (Eq.~\ref{eq:polytrope}, see also Fig.~\ref{fig:eos_mr_mrhat}) with indices $\gamma_1$, $\gamma_2$ , and $\gamma_3$ are attached, respectively ($n_0 < n_1 < n_2$). The step sizes used in the data generation were: $\delta_{\gamma_1} = \delta_{\gamma_3} = 0.25$, $\delta_{\gamma_2} = 0.1$, $\delta_{n_0} = 0.025$. In case of $\delta_{n_1}$ and $\delta_{n_2}$ the step varied during computation since the minimum values dependent on $n_0$ and $n_1$ , respectively,  but were not larger than $\delta_{n_0}$. }  
\label{tab:params}
\end{table}

For a given EOS, we solve the equations of hydrostatic equilibrium for a spherically
symmetric distribution of mass. The space-time metric is:   
\begin{equation} 
ds^2=e^{\nu(r)} c^2 dt^2 - \frac{dr^2}{1-2GM(r)/rc^2} - r^2(d\theta^2 + \sin^2\theta d\phi^2), \label{eq:tov-metric} 
\end{equation} 
with the gravitational mass $M(r)$ inside the radius $r$ 
\begin{equation} 
\frac{dM(r)}{dr}=4 \pi \rho(r) r^2. 
\label{eq:tov-dmdr}
\end{equation}
Then the resulting Tolman-Oppenheimer-Volkoff equations \citep{Tolman1939,OppenheimerV1939}, 
\begin{eqnarray} 
\frac{dP(r)}{dr}&=&-\frac{G}{r^2}\left(\rho(r)+\frac{P(r)}{c^2}\right)
\left(M(r)+\frac{4\pi r^3 P(r)}{c^2}\right) \nonumber \\
&\times& \left(1-\frac{2GM(r)}{c^2r}\right)^{-1},
\label{eq:tov-dpdr}
\end{eqnarray} 
supplied with the equation for the metric function $\nu(r)$, 
\begin{equation} 
\frac{d\nu(r)}{dr}= -\left(\frac{2}{P(r)+\rho(r)c^2} \right) \frac{dP(r)}{dr},  
\label{eq:tov-dnudr}
\end{equation} 
are integrated from the center towards the surface (where the pressure, $P,$
vanishes, which defines the radius of the star ,$R$) using a Runge-Kutta 4th order numerical scheme with a variable integration step \citep{Press1992} 
for a range of central parameters of the EOS (e.g. the
central pressures $P_c$) resulting in the $M(R)$ sequence. 

In addition to gravitational mass, $M,$ and radius, $R$, we also calculate the
static lowest-order tidal deformability of the star, defined as 
\begin{equation}
\lambda = \frac{2}{3}R^5k_2. 
\label{eq:lambda_small} 
\end{equation} 
The parameter $\lambda$ represents the star's reaction on the external tidal
field (e.g. exerted by a companion in a tight binary system, as observed in
\citealt{Abbott2017a}).  It is obtained in the lowest-order approximation, by
calculating the second (quadrupole) tidal Love number $k_2$ \citep{Love1911}, a
function of stellar parameters and hence the EOS:  
\begin{eqnarray} 
k_2 &= \frac{8}{5}x^5 (1-2x)^2 \bigl( 2-y + 2x(y-1)\bigr)   
\Bigl( 2x \bigl( 6 -3y +3x(5y-8)\bigr) \Bigr. \nonumber \\  
&+\ 4x^3 \left( 13 -11y + x(3y -2) + 2x^2(1+y) \right) \nonumber \\  
& \left.+\ 3(1-2x)^2 \bigl( 2 - y + 2x(y-1)\bigr) \ln(1-2x) \right)^{-1},  
\end{eqnarray} 
where $x=GM/Rc^2$ denotes the star's compactness, and $y$ the solution of 
\begin{eqnarray}
\frac{dy}{dr} = -\frac{y^2}{r}-\frac{1+4\pi G r^2/c^2(P/c^2-\rho)}{(r-2GM(r)/c^2)}y  \nonumber \\
+ \left(\frac{2G/c^2(M(r) + 4\pi r^3 P/c^2)}{\sqrt{r}(r-2GM(r)/c^2)}\right)^2 + \frac{6}{r-2GM(r)/c^2} \nonumber \\ 
- \frac{4\pi G r^2/c^2}{r-2GM(r)/c^2}\left(5\rho+ 9P/ c^2
+\frac{\left( \rho   + P/c^2\right)^2 c^2}{ \rho dP/d\rho}\right),
\end{eqnarray}
evaluated at the stellar surface \citep{FlanaganH2008,VanOeverenF2017}. In the
following we use the mass-normalised value of the $\lambda$ parameter, 
\begin{equation}
  \Lambda = \lambda\left( GM/c^2\right)^{-5}. 
  \label{eq:lambda} 
\end{equation} 
In order to relate the $\Lambda$ parameter with the stellar radius $R$, we produce a following radius-like parameter $\hat{R}(M, \Lambda)$, which we call the tidal radius (proposed in \citealt{Wade:2014vqa}): 
\begin{equation}
  \hat{R} = 2M\Lambda^{1/5}. 
  \label{eq:rhat} 
\end{equation} 
This function of $\Lambda$ and $M$ is henceforth used in the study. Sample $M(R)$ and $M(\hat{R})$ relations are presented in the top panels of Fig.~\ref{fig:eos_mr_mrhat}, along with simulated measurement points (the procedure of obtaining them is presented in Sect.~\ref{sec:mass_func}). Moreover, in the bottom right panel of Fig.~\ref{fig:eos_mr_mrhat}, we present a bundle of $M(R)$ relations used in the training, generated for piecewise relativistic polytropes to compare the training set with the astrophysical models based on the SLy4, APR and BSK20 EOSs. The training data cover the space of parameters similar to astrophysical models; therefore, we expect that the algorithm will generalise the EOS reconstruction toward previously unseen types of curves (types of curves on which it wasn't trained on). 

\subsection{Neutron-star mass function and simulated measurement errors} \label{sec:mass_func} 

In order to investigate the influence of the amount and precision of data - the number of observations $N$ and their measurement errors - we restrict the sample of $M(R)$ and $M(\hat{R})$ for masses from the astrophysically-realistic range above 1 $M_\odot$, which corresponds to observed NS masses in Galactic binary NS systems \citep{ns_masses}, and in the GW170817 and GW190425 events  \citep{2017PhRvL.119p1101A,2020arXiv200101761T}. We exclude from further analysis all piecewise polytropic solutions that are not compatible with current state of observations, giving the NS maximum masses above 1.9 $M_\odot$ (conservative choice motivated by the observations of massive NSs, see  \citealt{Demorest2010,Fonseca2016,Antoniadis2013,Cromantie2020}). 

 To realistically recreate astrophysical observations, we select the measurement points from a realistic NS mass function (mass probability distribution) out of which the mass values are to be randomly selected. Consistently with current observations of NSs in the Galaxy, the mass function is represented by a double-Gaussian distribution \citep{ns_masses} with the main peak around the Chandrasekhar mass and the second, smaller peak corresponding to the NS masses close to $2\ M_\odot$, namely $\mathcal{N}(\mu_1, \sigma_1) + \mathcal{N}(\mu_2, \sigma_2)$, where $\mu_1 = 1.34$, $\sigma_1 = 0.07$, $\mu_2 = 1.80$, $\sigma_2 = 0.21$ (see Fig.~1 in \citealt{ns_masses} for details). This NS mass function is consistent with a recent GW observation of a heavy NS binary system \citep{2020arXiv200101761T}.

The training data set is prepared by assuming that the measurements are witness to a measurement error. After randomly choosing $N$ values of the gravitational mass $M$ from the above-mentioned mass distribution, we construct the training samples corresponding to a given $M(R)$ or $M(\hat{R})$ point  by drawing values from normal distributions $\mathcal{N}(M(R), \sigma_i)$ or $\mathcal{N}(M(\hat{R}), \sigma_i)$, with $i=M, R, \hat{R}$ respectively. For the $\sigma_i$ parameters, we chose the values in the range of $0.01 - 0.1 M_\odot$ for $\sigma_M$ and $0.01 - 1$ km for $\sigma_R$. Uncertainties for tidal deformabilities are defined in terms of $\hat{R}$ and were in range $\sigma_{\hat{R}}=0.01-1$ km, which corresponds to $\sigma_{\Lambda}=10^2-10^3$.  An example of the training sequence, obtained assuming the double-Gaussian mass distribution and $\sigma_M = 0.1\ M_\odot$, $\sigma_R = \sigma_{\hat{R}} = 1.0$ km is shown in Fig.~\ref{fig:eos_mr_mrhat} (marked red in top panels). Gray curves (and the red curve) correspond to $M(R)$ and $M(\hat{R})$ relations computed with the TOV equations for some examples of the piecewise polytropic EOS described in Sect.~\ref{sec:eos_struct}. The scattered points correspond to the actual input data fed to our model; they are based on the red curve values according to the procedure describe above.

In total the training dataset contains 13982 piecewise polytrope EOSs (see Tab.~\ref{tab:params} for the details), out of which the $M(R)$ and $M(\hat{R})$ sequences were produced by solving the TOV equations. For each of these sequences, we then randomly selected $N$ values of $M$ ($N$ equal to 10, 15, 20, 30, 40, or 50 observations) using the above-mentioned NS mass distribution, and recover the corresponding values of $R$ and $\hat{R}$. For each input EOS, this procedure is repeated a fixed number of $N_s=30$ times. As a result, each input EOS is represented in the training stage by $N_s$ different realisations of $N$ observations of $M(R)$ or $M(\hat{R})$,  subject to `observational errors' by drawing the values from normal distributions parametrised by $\sigma_i$. This step allows us to effectively estimate the errors that ANN makes in the prediction of output sequences, that is, the error of reconstructing pressures and densities. To compute these errors, we then calculate the differences between the estimated output and the ideal expected result (the `ground truth' values). The errors are averaged for each measurement in a given collection of realisations. This step is repeated for all the EOSs in the training dataset, returning the set of error distributions: in the case of 20 measurements, we recover 20 distributions. The error bars presented for the output values in Sect.~\ref{sec:results} are the mean values of these distributions.

We contrast the reconstruction errors with the ANN loss function as they represent different features. Loss function is a metric defining overall performance of the ANN in terms of how well the predicted values are to original ground truth values in general. Reconstruction errors give detailed information about differences between predicted pressures and densities and their corresponding ground truth values. Furthermore, the reconstruction error changes with respect to the values of pressure and density.

At the last step of data preparation, pressures and densities were converted to the decimal logarithm values and scaled together with masses, radii, and tidal radii to the range $(0,1)$. Rescaling is required by the ANN non-linear functions since their domain is in the range $(0,1)$.

The data sets were then split into two separate subsets: a training set (70\% of all instances from the total dataset) and the testing set (30\% of the total dataset). In cases when the ANN was tested against the measurements corresponding to realistic tabulated EOS, the simulated measurement data was generated in the same way as for the piecewise polytropic EOS. 

\subsection{ANN} 
\label{ssec:ann_arch}

In the design of the ANN, we used parts of the AE architecture. The AE \citep{Kramer1991NonlinearPC} is a specific type of network capable of learning how to efficiently compress and encode the data into the so-called latent space representation and, later, to decompress and reconstruct the initial data as closely as possible. The core functionality of the AE is data dimensionality reduction. During training, AE learns how to ignore the noise and extract only crucial features of the data. Dimensionality reduction is in particular useful in the application of AEs aiming for data clustering. Specifically, the features of latent representation of an AE may be used to characterise the data, for example, by employing the conditional training of the variational AE using the training data with parameter labels to subsequently study the distribution of parameters in the latent space of variables. In the present exploratory work, we employ the simplest encoder-decoder structure of AE and we do not use the properties of the latent space, leaving that aspect to a future work. 

The final architecture of our ANN was chosen based on empirical tests based on the data. As an output criterion for the loss function we use the mean squared error (MSE). We tested architectures ranging from one to eight hidden layers. The optimal network, reaching the minimum value for MSE, was the one containing four hidden layers with the following number of neurons: 512, 256, 256, 512. 

The final set of hyper-parameters used for the training was the following (parameters defined as in e.g. \citealt{Goodfellow2016}): 

\begin{itemize} 
    \item {\tt ReLU} as the activation function for hidden layers
    \item sigmoid activation function for the output layer 
    \item ADAM optimiser \citep{ADAM2014} 
    \item batch size of 128 
    \item 0.001 learning rate
\end{itemize} 

The ANN architecture was implemented using the Python Keras library \citep{chollet2015keras} on top of the TensorFlow library \citep{tensorflow}, with support for the GPU. We developed the model on the NVidia Quadro P6000\footnote{Benefiting from the donation via the NVidia GPU seeding grant.} and performed the production runs on the Cyfronet Prometheus cluster\footnote{Prometheus, Academic Computer Centre CYFRONET AGH, Kraków, Poland} equipped with Tesla K40 GPUs, running CUDA 10.0 \citep{cuda} and the cuDNN 7.3.0 \citep{cuDNN}.

\section{Results}
\label{sec:results}

The results presented below are split into subsections. The first present the results of EOS reconstruction from $M(R)$ and $M(\hat{R})$ simulated measurements with errors using ANN trained on piecewise polytropic EOS results. The second subsection shows the application of ANN on the realistic EOS resulting from microscopic calculations (SLy4 EOS, \citealt{DouchinH2001}), that is, a reconstruction of the EOS which is not a piecewise polytropic model. We also study an application of the ANN to a direct reconstruction of the NS radius with the GW-only observations of the tidal deformability. 

\subsection{Translating the NS observations, $M(R)$ or $M(\Lambda)$, to EOS}
\label{ssec:mrrh_eos}

\begin{figure}
    \centering
    \includegraphics[width=0.45\textwidth]{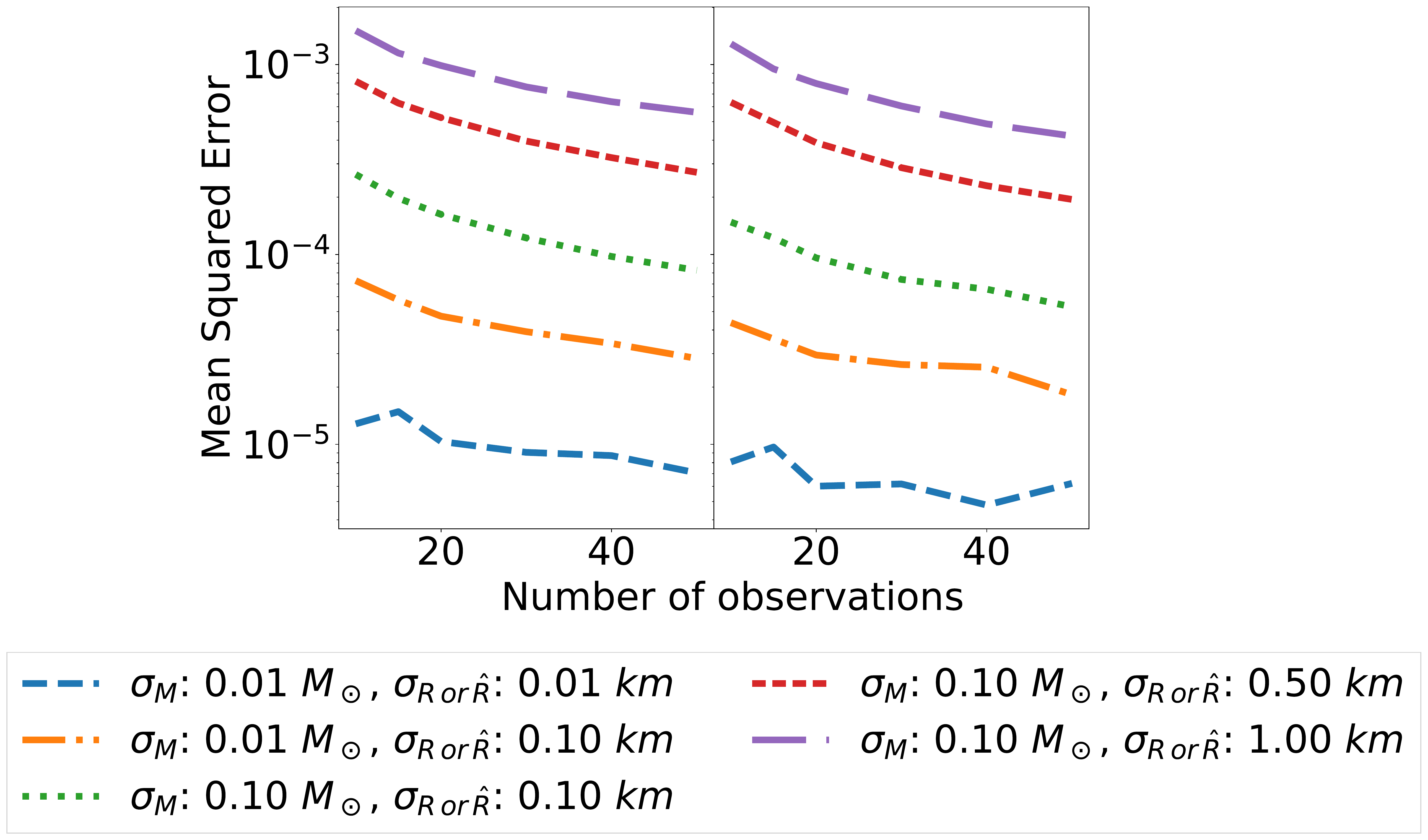}
    \caption{Dependence of the MSE (ANN loss function described in Sect.~\ref{sec:mass_func}) as a function of number of observations $N$ and measurement uncertainties in the case of $EOS$ estimation based on $M(R)$ (left figure) and $M(\hat{R})$ (right figure) observations.}
    \label{fig:mre_loss}
\end{figure}

Here, we present the results of the ANN application to the reconstruction of the EOS based on the gravitational mass, $M,$ and radius, $R,$ observations, which may be a result of electromagnetic observations of, for example, the NICER mission, as well as the EOS reconstruction based on the gravitational-wave observations of mass, $M,$ and tidal deformability, $\Lambda$ (which we reparametrise as $\hat{R}$; see Eq.~\ref{eq:rhat}). The ANN described in Sect.~\ref{ssec:ann_arch} is trained on data sets with varying number of observations and measurement uncertainties. The resulting figures of merit - the ANN loss function MSE - are shown in Fig.~\ref{fig:mre_loss} with the left plot corresponding to the EOS reconstruction using $M(R)$ data and the right using $M(\hat{R})$ data.

The accuracy of EOS estimation is mostly influenced by the assumed measurement uncertainties in both presented cases. The value of MSE is proportional to the measurement errors; it reaches the highest value for the largest of considered uncertainties: $\sigma_M = 0.1\ M_\odot$ for mass, $M$, $\sigma_R = 1$ km for the radius, $R,$ and $\sigma_{\hat{R}} = 1$ km for the tidal radius, $\hat{R}$. Furthermore, the number of observations $N$ had little effect on the MSE; the increase in $N$ slightly decreased the MSE in all studied cases.

\begin{figure}
    \centering
    \includegraphics[width=0.45\textwidth]{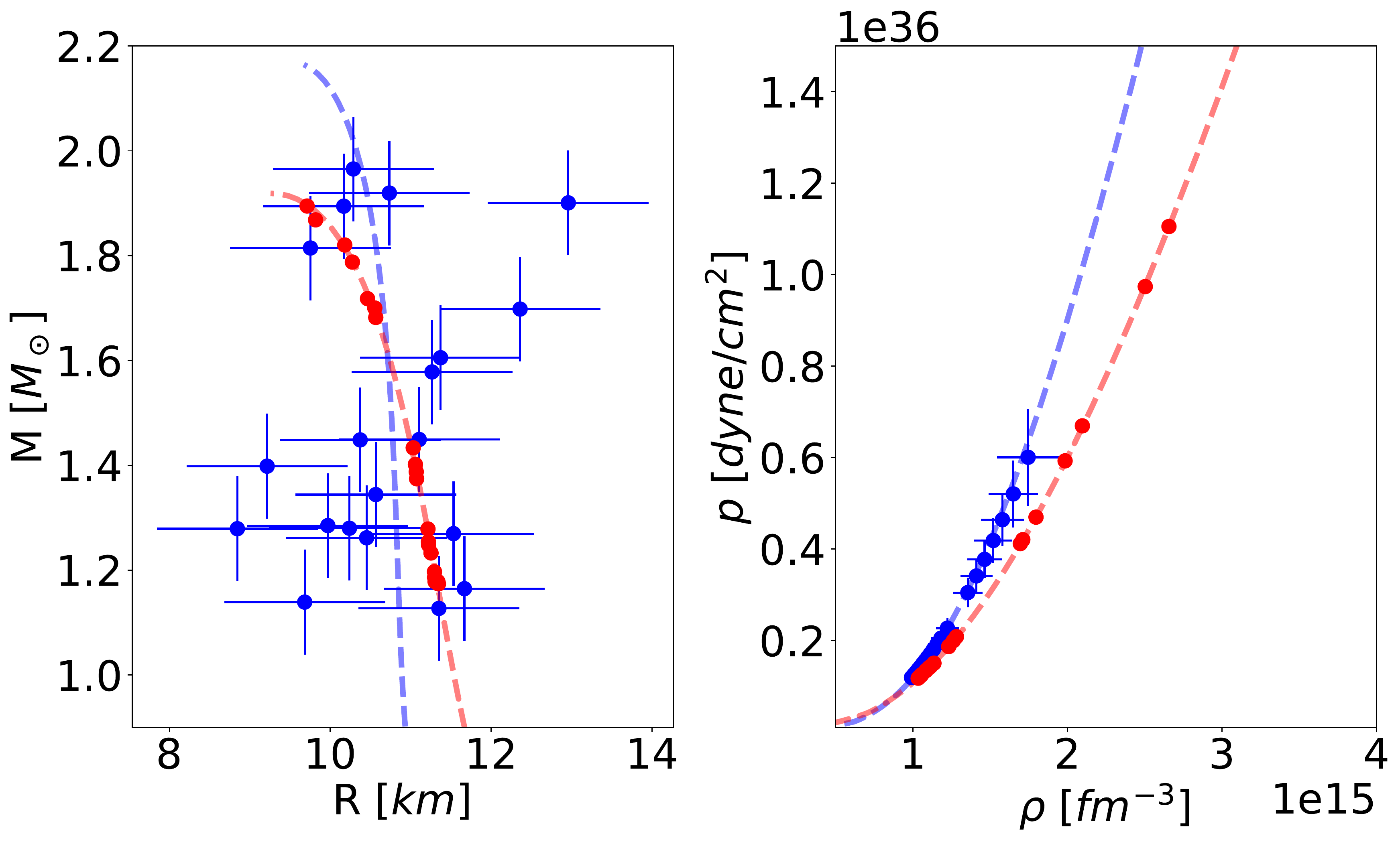}
    \vskip 5pt 
    \includegraphics[width=0.45\textwidth]{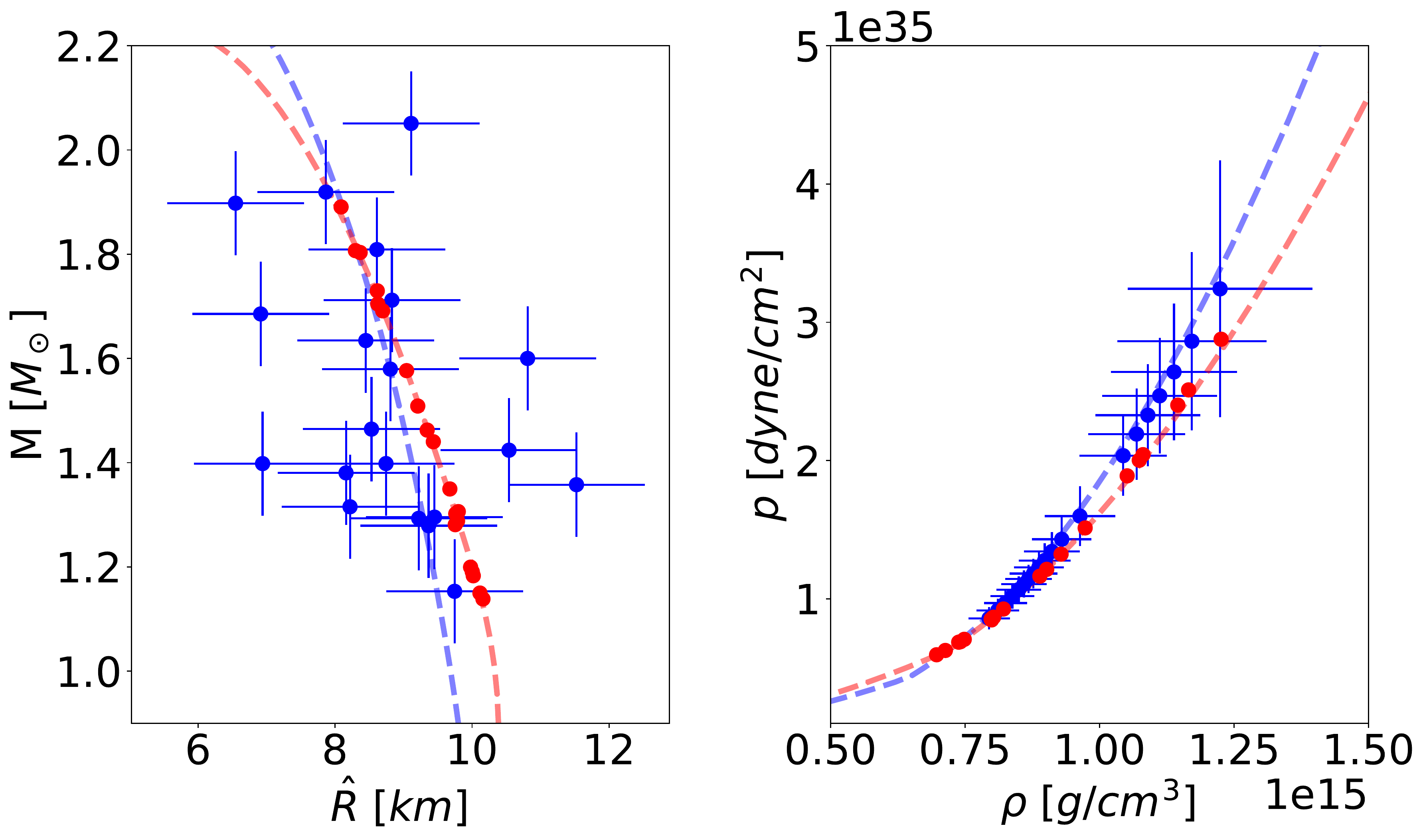}
    
    \caption{Top panels: Example of the input data ($M(R)$ measurements with errors, left plot), and corresponding output data from ANN ($p(\rho)$ relation, right plot) for the estimation of EOS from $M(R)$. Both input samples consist of 20 observations with masses randomly selected from a mass distribution (Sect.~\ref{sec:mass_func}) and measurement uncertainties equal to $\sigma_M=0.1\ M_\odot$, $\sigma_R=1$ km (blue sample), $\sigma_M=0.01\ M_\odot$, $\sigma_R=0.01$ km (red sample). For the description of the uncertainties on the output, see the text.  Bottom panels: Example of the input data ($M(\hat{R})$ measurements with errors, left plot), and corresponding output data from ANN ($p(\rho)$ relation, right plot) for the estimation of EOS from $M(\hat{R})$. Both input samples consist of 20 observations with masses randomly selected from a mass distribution (Sect.~\ref{sec:mass_func}) and measurement uncertainties equal to $\sigma_M=0.1\ M_\odot$, $\sigma_R=1$ km (blue sample), $\sigma_M=0.01\ M_\odot$, $\sigma_{\hat{R}}=0.01$ km (red sample). Dashed curves correspond to original (ground-truth, error-free) sequences of input and output of the TOV equations. Presented examples correspond to different EOSs.}
    \label{fig:mrrheos_results}
\end{figure}

The top panels of Fig.~\ref{fig:mrrheos_results} present two examples of the EOS reconstruction for the small and large measurement uncertainties in the case of $N=20$ $M(R)$  observations.  Both EOSs are recovered correctly within reconstruction errors computed as specified in the Sect.~\ref{sec:mass_func} with respect to the ground-truth values of related input EOSs (marked with dashed lines on the right panel). The error ranges in case of EOS estimation using $M(R)$ data for different measurement uncertainties are presented in the upper part of Table~\ref{tab:rec_errors}. The resulting $\sigma_p$ and $\sigma_{\rho}$ spans increase proportionally, with increasing $\sigma_M$ and $\sigma_R$. Furthermore, in all presented cases, the ranges for pressure errors were wider than density errors, indicating that ANN was more uncertain in the reconstruction of pressure values. The increase in the reconstruction errors is expected because the overall performance of the ANN was worse during the training (see the blue and violet curves in Fig.~\ref{fig:mre_loss} for comparison). Another effect related to the worse performance of the EOS reconstruction is the significant increase of errors and decrease in the accuracy of reconstruction for higher $p(\rho)$ values. Several effects may be responsible for this result, for example, the impact of adopted NS mass distribution. Naturally, if the dataset contains a smaller number of high (close to 2 $M_\odot$) $M$ samples, the high $p(\rho)$ values of the EOS are less efficiently probed. As a result, the EOS reconstruction is  less certain overall in this range. We discuss alternative explanations in Sect.~\ref{sec:discussion}.

\begin{table}
  \centering
  \resizebox{\columnwidth}{!}{
  \begin{tabular}{| c | c | c |} 
    \hline 
    $M(R)$ input data & $\sigma_p$ [dyne/cm$^2$] & $\sigma_{\rho}$ [g/cm$^3$]
    \\ \hline\hline 
    $\sigma_M=0.01\,M_\odot$, $\sigma_R=0.01$ km & $5\cdot10^{12}-10^{13}$ & $10^{33} - 7\cdot10^{33}$ \\ \hline
    $\sigma_M=0.01\,M_\odot$, $\sigma_R=0.1$ km & $9\cdot10^{12}-4\cdot10^{13}$ & $2\cdot10^{33} - 2\cdot10^{34}$ \\ \hline 
    $\sigma_M=0.1\,M_\odot$, $\sigma_R=0.1$ km & $10^{13}-5\cdot10^{13}$ & $5\cdot10^{33} - 4\cdot10^{34}$ \\ \hline 
    $\sigma_M=0.1\,M_\odot$, $\sigma_R=0.5$ km & $2\cdot10^{13}-10^{14}$ & $5\cdot10^{33} - 8\cdot10^{34}$ \\ \hline 
    $\sigma_M=0.1\,M_\odot$, $\sigma_R=1.0$ km & $4\cdot10^{13}-2\cdot10^{14}$ & $8\cdot10^{33} - 10^{35}$ \\ \hline 
  \end{tabular}
  }
\newline
\vspace*{0.2cm}
\newline
  \centering 
  \resizebox{\columnwidth}{!}{
  \begin{tabular}{| c | c | c |} 
    \hline 
    $M(\hat{R})$ input data & $\sigma_p$ [dyne/cm$^2$] & $\sigma_{\rho}$ [g/cm$^3$]
    \\ \hline\hline 
    $\sigma_M=0.01\,M_\odot$, $\sigma_{\hat{R}}=0.01$ km & $5\cdot10^{12}-8\cdot10^{12}$ & $10^{33} - 4\cdot10^{33}$ \\ \hline
    $\sigma_M=0.01\,M_\odot$, $\sigma_{\hat{R}}=0.1$ km & $9\cdot10^{12}-3\cdot10^{13}$ & $2\cdot10^{33} - 2\cdot10^{34}$ \\ \hline 
    $\sigma_M=0.1\,M_\odot$, $\sigma_{\hat{R}}=0.1$ km & $10^{13}-4\cdot10^{13}$ & $4\cdot10^{33} - 2\cdot10^{34}$ \\ \hline 
    $\sigma_M=0.1\,M_\odot$, $\sigma_{\hat{R}}=0.5$ km & $2\cdot10^{13}-10^{14}$ & $5\cdot10^{33} - 7\cdot10^{34}$ \\ \hline 
    $\sigma_M=0.1\,M_\odot$, $\sigma_{\hat{R}}=1.0$ km & $4\cdot10^{13}-2\cdot10^{14}$ & $8\cdot10^{33} - 9\cdot10^{34}$ \\ \hline 
  \end{tabular}
  }
\vskip 1em  
\caption{Reconstruction error ranges for $\sigma_{p}$ and $\sigma_{\rho}$ of the ANN for studied measurement uncertainties in case of EOS reconstruction for the $M(R)$ data (upper table) and $M(\hat{R})$ data (lower table). The reconstruction errors are computed as specified in Sect. \ref{sec:mass_func}.}
\label{tab:rec_errors}
\end{table}

The examples shown in the bottom panels of Fig.~\ref{fig:mrrheos_results} corresponded to the EOS reconstruction using $M(\hat{R})$ data for two cases of small and large measurement uncertainties and $N=20$ observations. Both EOSs are estimated correctly within reconstruction errors with respect to the ground-truth values of corresponding EOSs (marked with dashed lines) and the errors are proportional to the values of the density and pressure, similarly to the $M(R)$ case.

\subsection{Application on realistic EOS}
\label{ssec:sly4}

\begin{figure}
    \centering
    \includegraphics[width=0.45\textwidth]{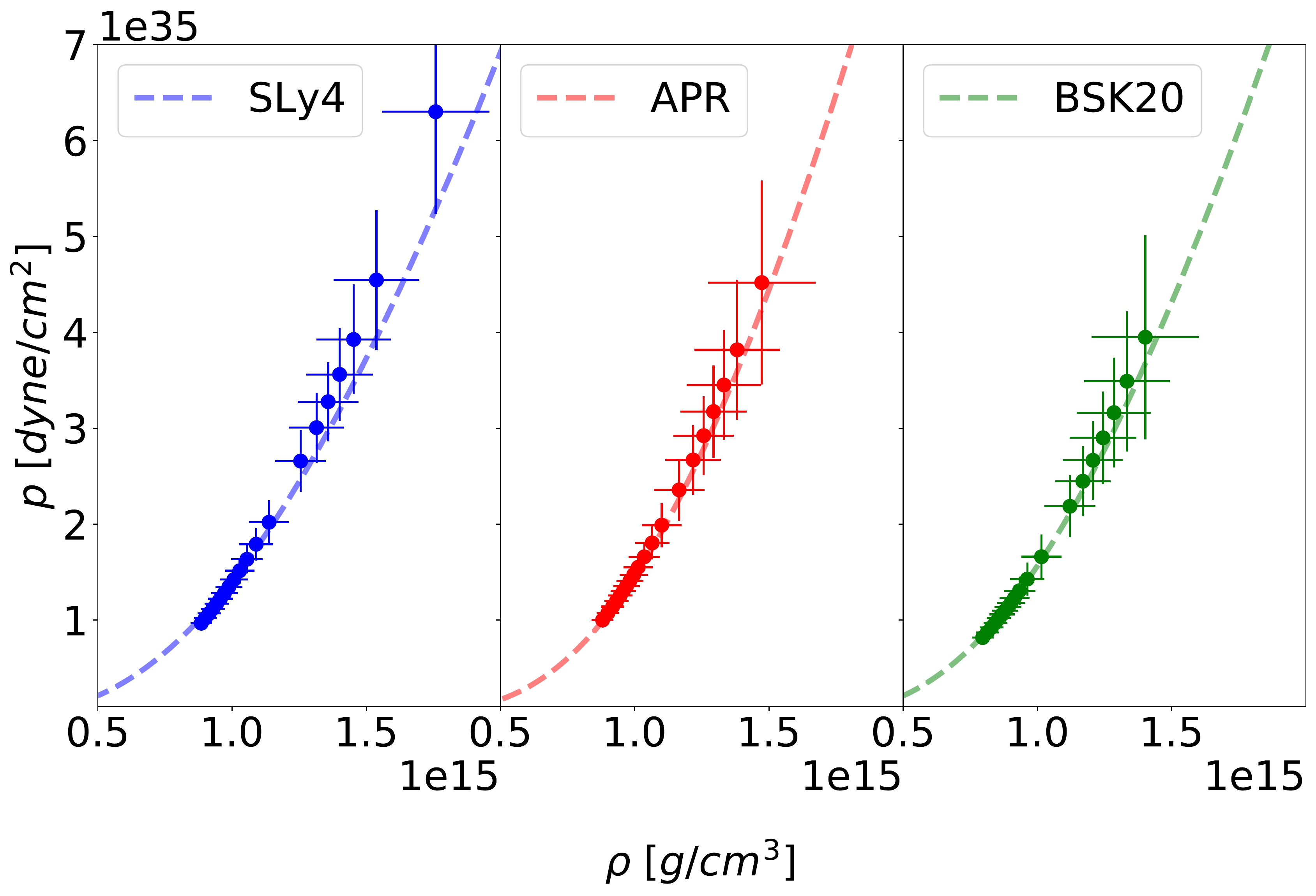}
    \vskip 5pt 
    \includegraphics[width=0.45\textwidth]{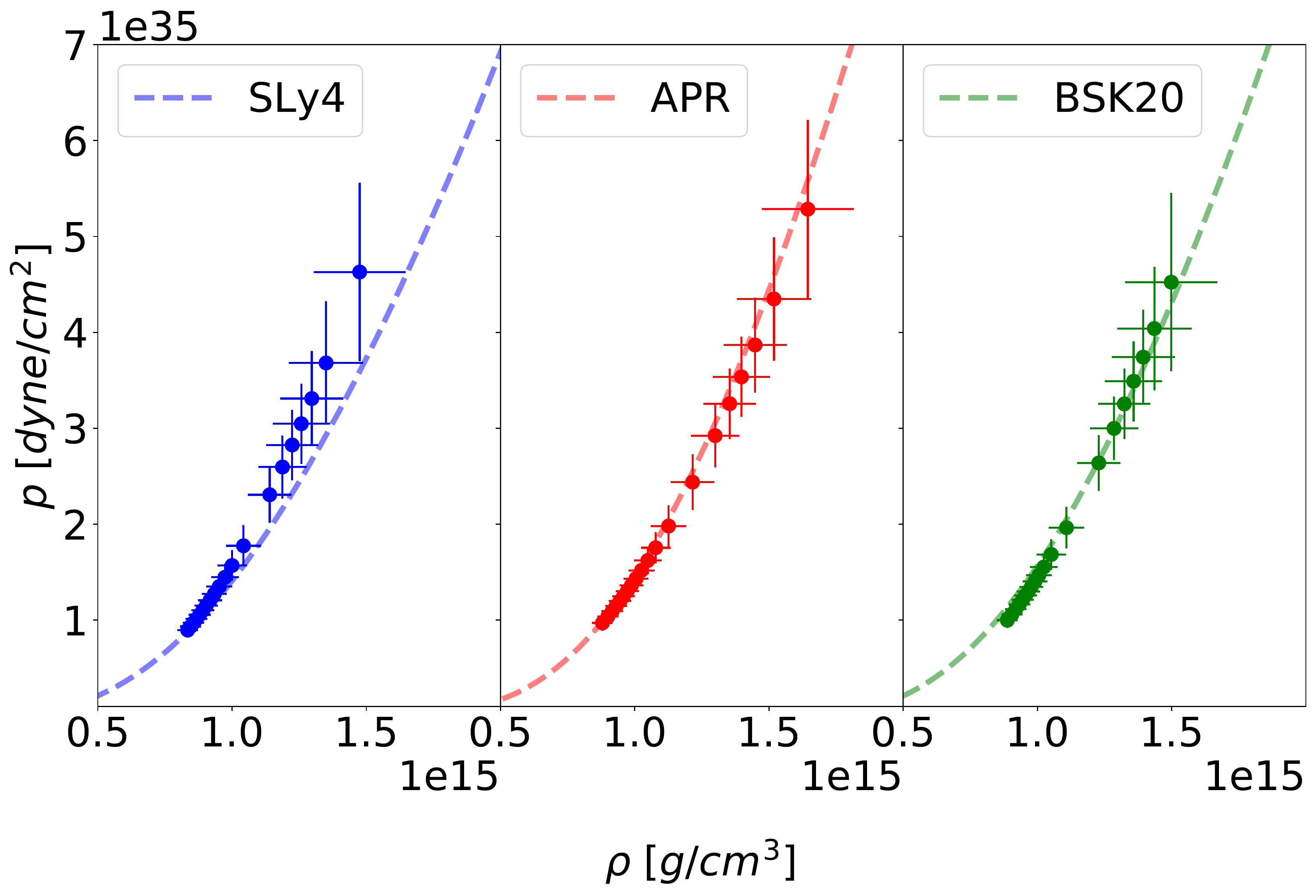}
    \caption{Top panels: ANN-reconstructed EOS from the $M(R)$ data for the SLy4 EOS model (left plot), the APR EOS model (middle plot) and the BSK20 EOS model (right plot). Results are computed for the input $M(R)$ data consisting of 20 observations with measurement uncertainties equal to $m_{err}=0.1 M_\odot$, $\hat{r}_{err}=1$ km.
    Bottom panels: ANN-reconstructed EOS from the $M(\hat{R})$ data for the same EOS as in the top panels. Results for the input $M(\hat{R})$ data consisting of 20 observations with measurement uncertainties equal to $m_{err}=0.1 M_\odot$, $\hat{r}_{err}=1$ km. Dashed lines correspond to the exact EOS relations.}
    \label{fig:mrrheos_real_results}
\end{figure}

We test the ANN trained on piecewise polytropic EOS (and the TOV solutions obtained with them) on a realistic microscopic EOSs: the SLy4 EOS \citep{DouchinH2001}, the APR EOS \citep{PhysRevC.58.1804} and the BSK20 EOS \citep{PhysRevC.82.035804}. To generate data for this test, we followed the approach detailed in Sect.~\ref{sec:prep} as in the case of the polytropic EOSs. Figure~\ref{fig:mrrheos_real_results} contain the results of EOS reconstruction using the $M(R)$ data (top panels) and $M(\hat{R})$ (bottom panels) for $N=20$, $\sigma_M=0.1\,M_\odot$ and $\sigma_R=1\,km$. Among the realistic microphysical EOS we have considered, the EOS relation reconstructed for the APR EOS and BSK20 EOS agree with original (ground truth) input values almost perfectly, whereas the SLy4 EOS model is reconstructed less precisely; however, the reconstructed EOS relation agrees with the ground truth values (dashed line) within reconstruction errors from Table~\ref{tab:rec_errors}.

These results show that the ANN trained on a relatively simple dataset of relativistic piecewise polytropes is able to generalise the task of EOS reconstruction towards an unknown during its training of realistic EOS.

\subsection{Radius reconstruction using $\Lambda$ measurements}
\label{ssec:lambdar}

We also present the results of an additional analysis which aims to directly reconstruct the NS radius $R$ from GW-only observations of masses and tidal deformabilities. As Eq.~\ref{eq:lambda} shows, the tidal deformability is related to $M$ and $R$ and to the second Love number $k_2$, all of which are functionals on the EOS. In the general case, the $\Lambda-R$ relation cannot be simply obtained (see e.g. \citealt{2019arXiv190902274Z,2018PhRvL.121i1102D} and references therein). From the point of view of the $M(R)$ diagram, the relation between $\Lambda$ and $R$ depends on the slope of $M(R)$, which is indirectly a function of the NS susceptibility to deformations (see \citealt{2019A&A...622A.174S} for examples of configurations with the same $M$ and $R$, but different $\Lambda$ values; their Sect. 3.2, Figs. 9 and 10). 

\begin{figure}
    \centering
    \includegraphics[width=0.45\textwidth]{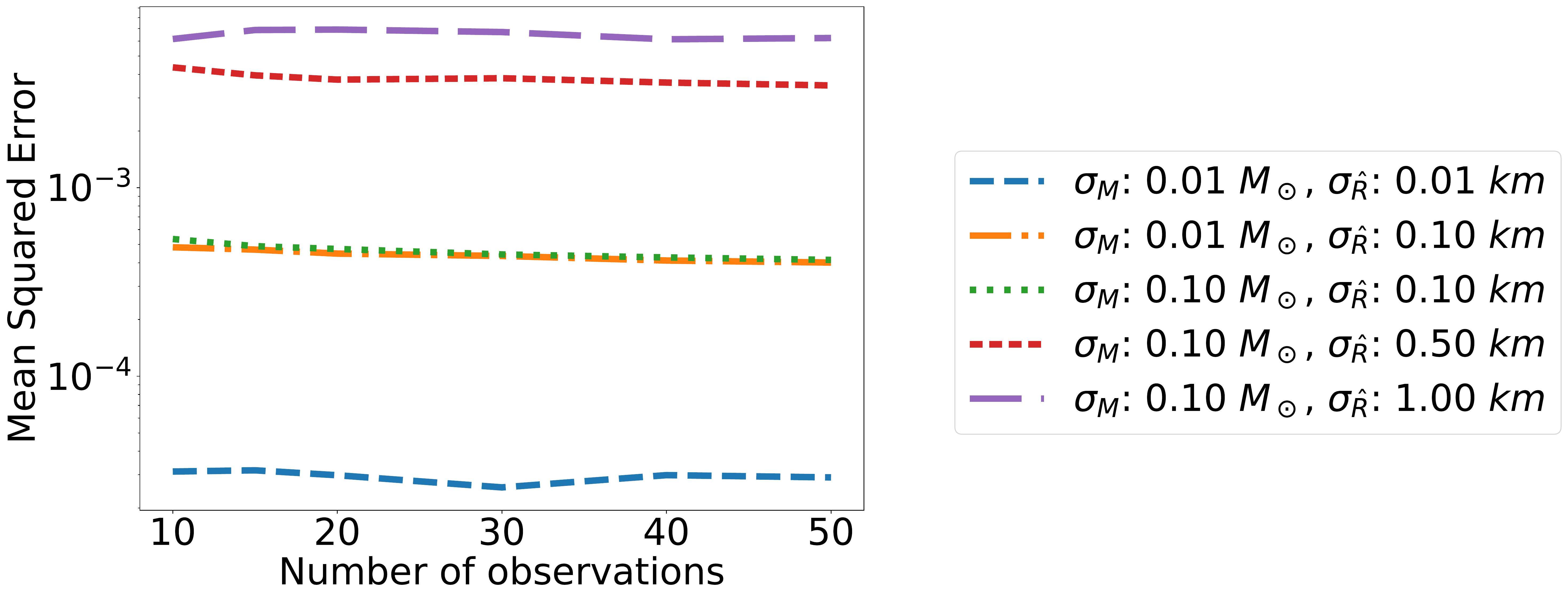}
    \caption{Evolution of MSE (ANN loss) in the function of number of observations and measurement uncertainties in case of $R$ computation based on $M(\hat{R})$. }
    \label{fig:mrrh_loss}
\end{figure}

In order to study the ability of reconstructing the $R$ based on $M$ and $\Lambda$ observations, we modified the ANN described in the Sect.~\ref{ssec:ann_arch} since, for this case, the size of the output was twice smaller ($M$ and $\hat{R}$ concatenated at the input and $R$ and at the output). We considered the same measurement uncertainties $\sigma_M$ and $\sigma_{\hat{R}}$ as in the EOS reconstruction. The results of the ANN training are shown in Fig. \ref{fig:mrrh_loss} in terms of MSE.

Similarly as in the estimation of EOS, the strongest influence on the radius computation had the measurement uncertainties. The MSE changed in range between $5\cdot10^{-5}$ and $10^{-2}$ for the data varying in uncertainties from $\sigma_M=0.01\ M_\odot$ and $\sigma_{\hat{R}}=0.01$ km to $\sigma_M=0.1\ M_\odot$ and $\sigma_{\hat{R}}=1$ km. Moreover, the impact of observations number was insignificant.

\begin{figure}
\centering
    \includegraphics[width=0.45\textwidth]{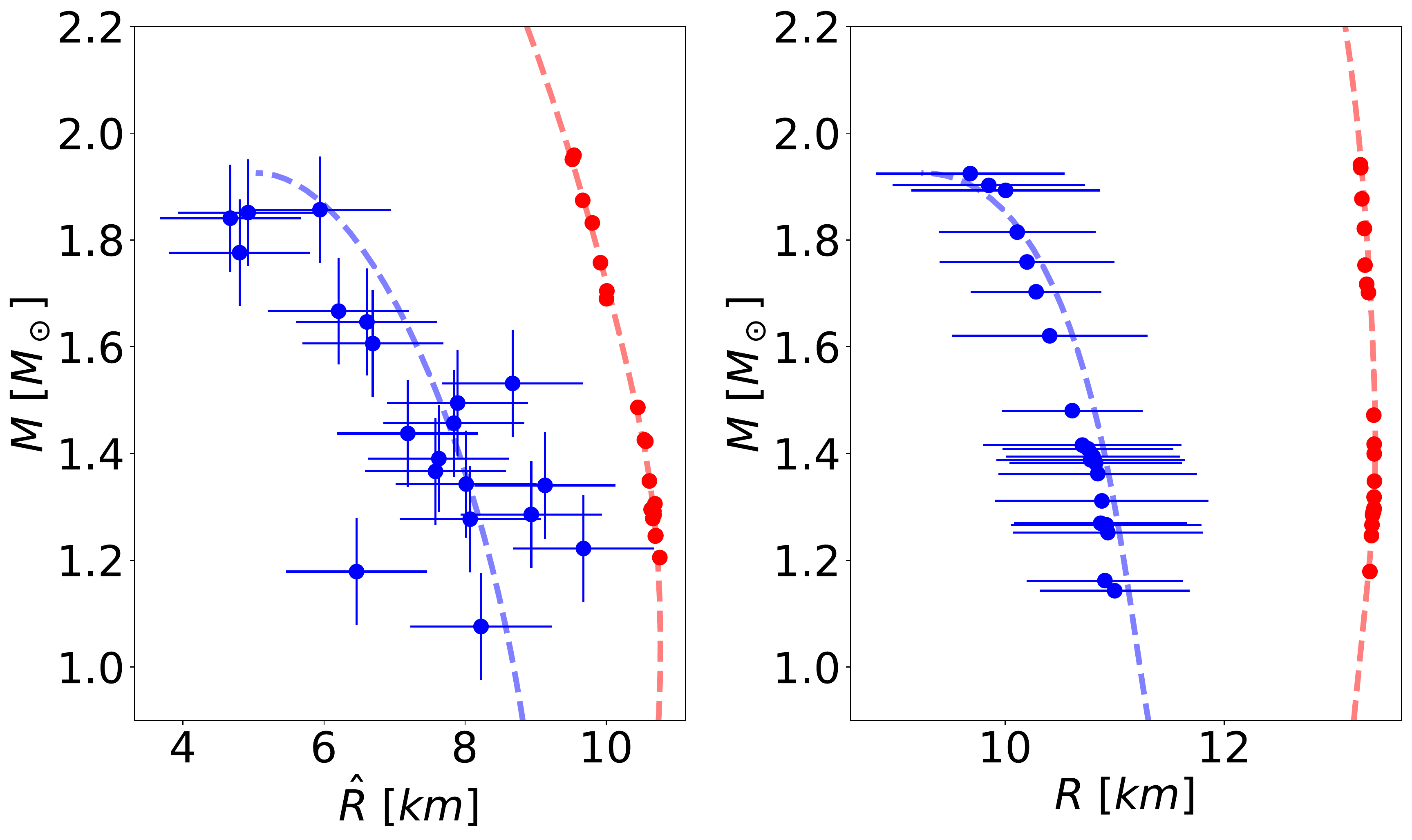}
    \vskip 5pt 
    \includegraphics[width=0.45\textwidth]{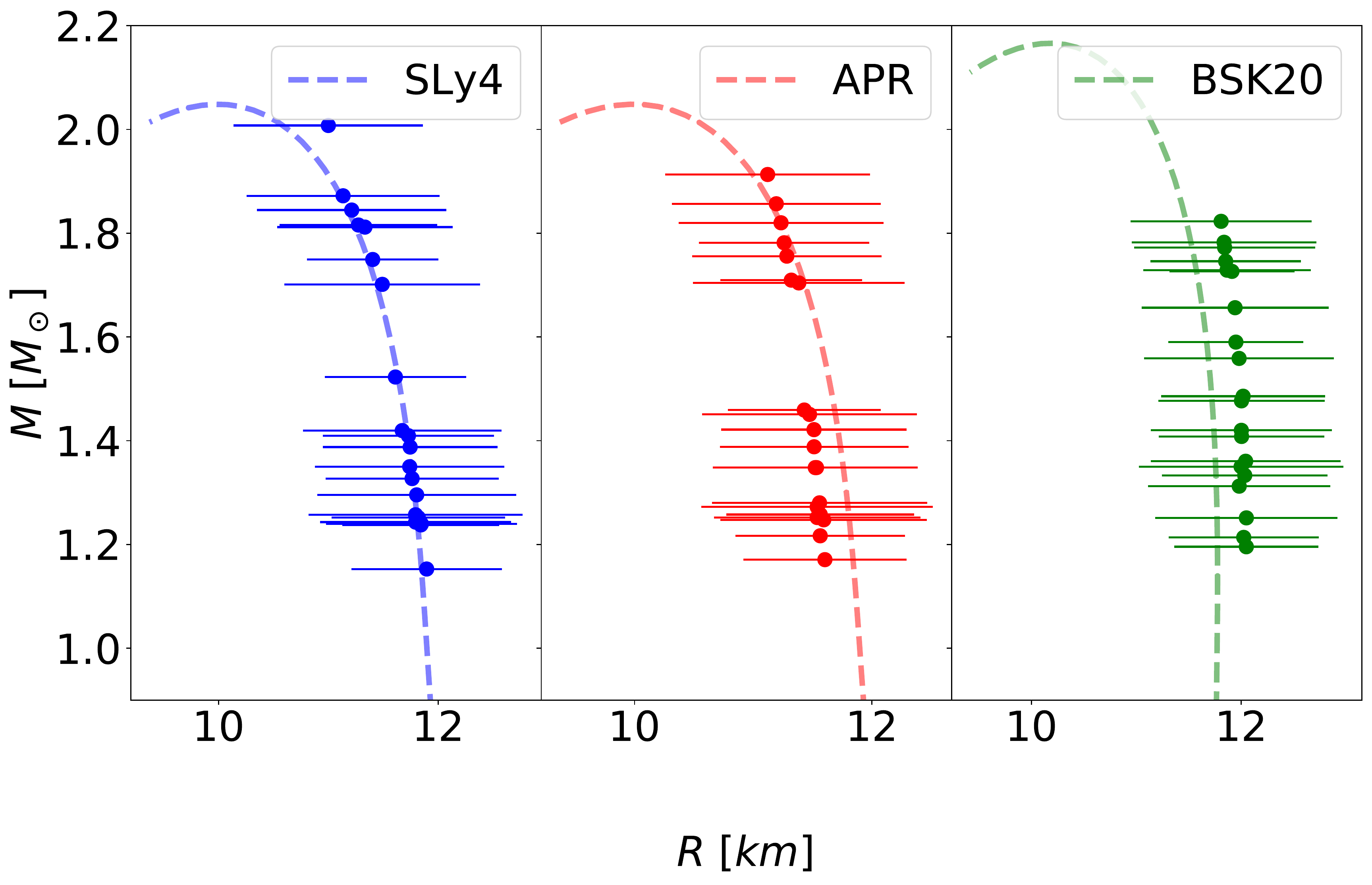}
    \caption{Top panels: Example of the input data (left plot) and corresponding output data from ANN (right plot) in the case of $R$ computation based on $M(\hat{R})$ for the piecewise relativistic polytropes. Bottom panels: Radius reconstructed from the $M(\hat{R})$ data for the SLy4 EOS model (left plot), the APR EOS model (middle plot), and the BSK20 EOS model (right plot).
    All results were computed for the input $M(R)$ data consisting of 20 observations with measurement uncertainties equal to $\sigma_M=0.1 M_\odot$, $\sigma_{\hat{R}}=1$ km. Dashed lines correspond to exact values obtained by solving the TOV equations.
    }
    \label{fig:mrrh_results}
\end{figure}

The examples of the radius estimation are presented in Fig.~\ref{fig:mrrh_results}. Top panels show the radius computed by ANN using piecewise polytropic data for two cases of measurement uncertainties: $\sigma_M=0.01\,M_\odot$ and $\sigma_{\hat{R}}=0.01$ km (red sample) and $\sigma_M=0.1\,M_\odot$ and $\sigma_{\hat{R}}=1.0$ km (blue sample).
Bottom panels present the estimated radius for the data corresponding to the realistic cases: the SLy4 EOS, APR EOS, and BSK20 EOS for $\sigma_M=0.1\,M_\odot$ and $\sigma_{\hat{R}}=1.0$ km. Within the reconstruction errors, $\sigma_R$ , all cases were correctly reconstructed, in comparison to the dash line representing the exact values of radii computed from the TOV equations. However, the $\sigma_R$ increase proportionally to $\sigma_M$ and $\sigma_{\hat{R}}$. Furthermore, the errors varied randomly with respect to the value of radius. In contrast to pressure and density errors, no trend in radius errors was present.

In general, the radius reconstruction from tidal deformability using ANN is possible, which demonstrates an additional ability on the part of ANN to build a non-linear mapping between astrophysical parameters of interest.

\section{Discussion}
\label{sec:discussion}

The above results point us to a conclusion that the application of ANN in EOS reconstruction from astrophysical observations works for the majority of our data, with decreasing reliability for data with the largest measurement errors. 

In comparison with similar approaches to the same problem \citep{2018PhRvD..98b3019F,PhysRevD.101.054016}, our work extends the study of applications of the NN to NS multi-messenger astrophysics in several ways: we directly output the $p(\rho)$ EOS table, not limiting the output to selected EOS parameters, meaning that our implementation is, in principle, not bound to specific prescription of the EOS. In addition, we study the application of the AE architecture to the problem of EOS reconstruction, investigate as input the tidal deformability parameters as a function of mass, not only $M(R)$, that is, we try to simulate a situation in which the data comes exclusively from GW measurements, and we also investigate varying number of measurements, measurement errors, and realistic mass functions; for an additional investigation related to the last point, see the text below. 

Motivated by the issue behind the significant increase of the reconstruction EOS errors for higher densities and pressures in cases of large measurement uncertainties ($m_{err}=0.1\ M_\odot$ and $r_{err} = 1$ km) we performed an additional analysis. We understand this as a feature of the non-linearity of the mapping between the observed values of $M$, $R$ and $\hat{R}$ and the EOS. As shown in Fig.~\ref{fig:mrrheos_results}, for example, the measurements at high masses probe a significantly larger range of pressure and densities than those at lower masses. In addition, the values of radii $R$ and tidal deformabilities $\Lambda$ (and hence $\hat{R}$) are typically smaller for larger masses: stars are more compact and also less prone to deformation. Sampling the measurements from the high-mass range, where the differences between measurements are small but the errors are comparable to the low-mass measurements, should result in worse reconstruction in the high pressure and density range of the EOS.

\begin{figure}
    \centering
    \includegraphics[width=0.45\textwidth]{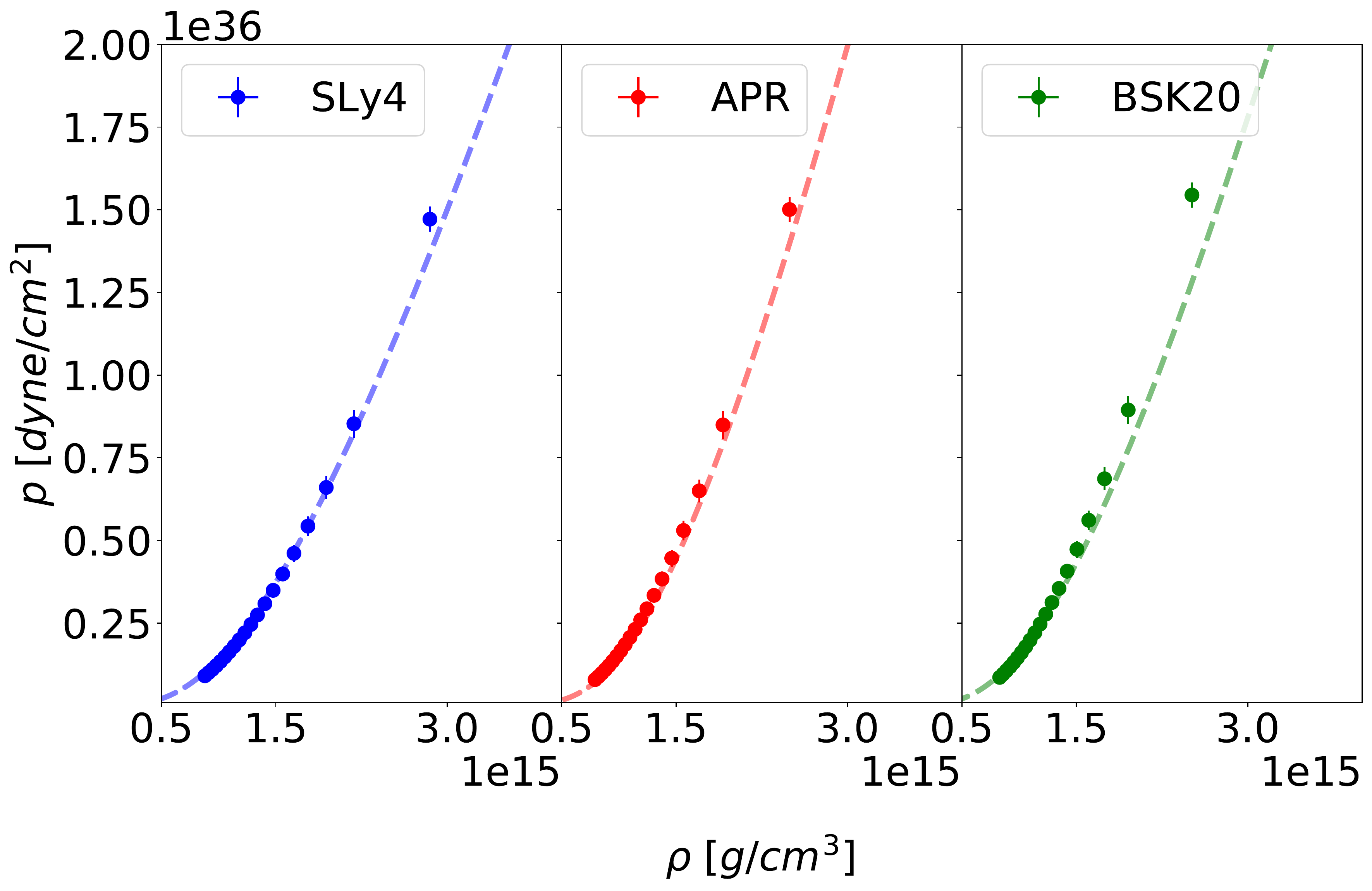}
    \caption{ANN reconstructed EOS using the $M(R)$ data for the SLy4 EOS model (left plot), the APR EOS model (middle plot) and the BSK20 EOS model (right plot) for the uniform NS mass distribution. The presented results were computed for input data consisting of 20 observations with measurement uncertainties equal to $\sigma_M=0.1 M_\odot$, $\sigma_R=1$ km. Dashed lines correspond to exact values obtained by solving the TOV equations.}
    \label{fig:mr_eos_results_distr}
\end{figure}

In order to study this further also from the point of view of the choice of mass function, we  performed additional simulations. Since the  double-Gaussian function we initially
adopted has its main distribution peak in the low-mass range (around the Chandrasekhar mass), the majority of generated observation points correspond to lower pressures and densities, which are precisely reconstructed by the algorithm. However, the high mass, and therefore the high pressure and density range, is covered sparsely; hence the corresponding high pressures and densities may be reconstructed less precisely. To test this explanation, a new training data using alternative NS mass distribution were prepared. We considered a uniform mass distribution in the range between 1 and 2.2 $M_\odot$. During the training on the uniform mass distribution data set, the ANN reached lower values of MSE with respect to the results presented in Sect. \ref{sec:results} with differences of around one order of magnitude in all considered cases. As a result the EOS reconstruction was characterised by smaller reconstruction errors for pressure and density; see examples of reconstruction in Fig. \ref{fig:mr_eos_results_distr} for 20 observations with $\sigma_M=0.1\,M_\odot$ and $\sigma_R=1\,km$. Moreover, predicted values probed range of higher values with respect to results from Sec. \ref{ssec:sly4}. The uniform mass distribution allow to generate observations close to maximum value of 2.2 $M_\odot$ (including measurement uncertainties), whereas the previously used double-Gaussian function returned masses rarely higher than 2 $M_\odot$. 

Our results suggest that to efficiently probe the high-mass end of the NS distribution, either measurement uncertainties should be significantly decreased with respect to the low-mass range or coverage of masses should be more uniform. The first possibility may be feasible with the 3rd generation GW detectors, such as the Einstein Telescope \citep{2020JCAP...03..050M}. On the other hand, the EOS is accurately reconstructed for the low-mass range (low pressure and density regime), which offers the possibility of comparing nuclear parameters with the data from terrestrial experiments. 

It is also worth mentioning that a precise reconstruction of EOS using ANN requires training data that is representative of the problem. In order to reconstruct astrophysical EOS models (SLy4, APR, and BSK20), we have selected an appropriate training set. However, ANN tested on different EOS covering different ranges for $M$, $R$, $\Lambda$, $p,$ and $\rho$ would result in a worse reconstruction. To avoid this problem, it's necessary to optimise the parameter space of the training set and choose astrophysical models accordingly. It would be straightforward to expand the training dataset with a specific parametric description of dense matter, such as the MIT bag, to describe the de-confined quark matter (\citealt{Chodos1974}, see \citealt{2019A&A...622A.174S} for an example of piecewise relativistic polytrope EOS supplemented by quark EOS approximation of \citealt{Zdunik2000}). In such a case, the ANN would potentially serve as a tool to discover the presence of exotic phases or signatures of dense-matter phase transitions.

\section{Conclusions}
\label{sec:summary}

We show that the ANN can be successfully applied in the reconstruction of the dense matter EOS from NS observations, either electromagnetic (masses and radii) or based on gravitational-wave measurements (masses and tidal deformabilities). We study the influence of the number of observations and the measurement uncertainties on the EOS reconstruction. The latter factor turned out to have a more significant effect on ANN performance, quantified in terms of the loss function (MSE). Furthermore, we show that the ANN trained on piecewise relativistic polytropes is capable of generalising the EOS reconstruction toward samples it wasn't previously exposed to: realistic EOSs resulting from microscopic calculations: the SLy4, APR, and BSK20 EOS models.

We also introduce reconstruction errors for ANN: $\sigma_{\rho}$ and $\sigma_p$. The presented values vary proportionally to either the uncertainties of measurement with regard to the observables or to the values of pressures and densities. To decrease reconstruction errors, we suggest that either measurement uncertainties should be reduced, which is possible with the new generation of telescopes and detectors (i.e. Einstein Telescope for gravitational observations), or masses should be generated more uniformly. Moreover, we show that ANN can be successfully used in the reconstruction of radius based on the gravitational observables, which can be particularly useful for gravitational astronomy.

Among the many possibilities for further development in studies of NS parameters using ML methods, we plan to focus on the promising direction of variational auto-encoders. The latent space of these algorithms contain features that allow for an in-depth understanding of the distribution of parameters of the input data. Studies of the latent space could be used, for example, to infer information on the nuclear parameters of the EOS or assess the plausibility of the existence of a dense-matter phase transition.

\begin{acknowledgements}
\label{sec:ack}

The work was partially supported by the Polish National Science Centre grant no. 2016/22/E/ST9/00037 and the European Cooperation in Science and Technology COST action G2net no. CA17137. The Quadro P6000 used in this research was donated by the NVIDIA Corporation. This research was supported in part by PL-Grid Infrastructure (the Prometheus cluster).  

\end{acknowledgements}

\bibliographystyle{aa}
\bibliography{bibfile}

\begin{thebibliography}{52}
\expandafter\ifx\csname natexlab\endcsname\relax\def\natexlab#1{#1}\fi

\bibitem[{Abadi {et~al.}(2015)Abadi, Agarwal, Barham, Brevdo, Chen, Citro,
  Corrado, Davis, Dean, Devin, Ghemawat, Goodfellow, Harp, Irving, Isard, Jia,
  Jozefowicz, Kaiser, Kudlur, Levenberg, Man\'{e}, Monga, Moore, Murray, Olah,
  Schuster, Shlens, Steiner, Sutskever, Talwar, Tucker, Vanhoucke, Vasudevan,
  Vi\'{e}gas, Vinyals, Warden, Wattenberg, Wicke, Yu, \& Zheng}]{tensorflow}
Abadi, M., Agarwal, A., Barham, P., {et~al.} 2015, {TensorFlow}: Large-Scale
  Machine Learning on Heterogeneous Systems, software available from
  tensorflow.org

\bibitem[{{Abbott} {et~al.}(2020){Abbott}, {Abbott}, {Abbott}, {Abraham},
  {Acernese}, {Ackley}, {Adams}, {Adhikari}, {et~al.}}]{2020arXiv200101761T}
{Abbott}, B.~P., {Abbott}, R., {Abbott}, T.~D., {et~al.} 2020, arXiv e-prints,
  arXiv:2001.01761

\bibitem[{{Abbott} {et~al.}(2017{\natexlab{a}}){Abbott}, {Abbott}, {Abbott},
  {Acernese}, {Ackley}, {Adams}, {Adams}, {Addesso}, {Adhikari}, {Adya},
  {Affeldt}, {Afrough}, {Agarwal}, {Agathos}, {Agatsuma}, {Aggarwal}, {Aguiar},
  {Aiello}, {Ain}, {Ajith}, {Allen}, {Allen}, {Allocca}, {Altin}, {Amato},
  {Ananyeva}, {et~al.}}]{2017PhRvL.119p1101A}
{Abbott}, B.~P., {Abbott}, R., {Abbott}, T.~D., {et~al.} 2017{\natexlab{a}},
  \prl, 119, 161101

\bibitem[{Abbott {et~al.}(2018)Abbott, Abbott, Abbott, Acernese, Ackley, Adams,
  Adams, Addesso, Adhikari, Adya, Affeldt, Agarwal, Agathos,
  {et~al.}}]{PhysRevLett.121.161101}
Abbott, B.~P., Abbott, R., Abbott, T.~D., {et~al.} 2018, Phys. Rev. Lett., 121,
  161101

\bibitem[{{Abbott} {et~al.}(2017{\natexlab{b}}){Abbott}, {Abbott}, {Abbott},
  {Acernese}, {Ackley}, {Adams}, {Adams}, {Addesso}, {Adhikari}, {Adya},
  {Affeldt}, {et~al.}}]{2017ApJ...848L..13A}
{Abbott}, B.~P., {Abbott}, R., {Abbott}, T.~D., {et~al.} 2017{\natexlab{b}},
  \apjl, 848, L13

\bibitem[{Abbott {et~al.}(2019)Abbott, Abbott, Abbott, Acernese, Ackley, Adams,
  Adams, Addesso, Adhikari, Adya, {et~al.}}]{PhysRevX.9.011001}
Abbott, B.~P., Abbott, R., Abbott, T.~D., {et~al.} 2019, Phys. Rev. X, 9,
  011001

\bibitem[{Abbott {et~al.}(2017)Abbott, Abbott, Abbott, Acernese, Ackley,
  {et~al.}}]{Abbott2017a}
Abbott, B.~P., Abbott, R., Abbott, T.~D., {et~al.} 2017, Phys. Rev. Lett., 119,
  161101

\bibitem[{Akmal {et~al.}(1998)Akmal, Pandharipande, \&
  Ravenhall}]{PhysRevC.58.1804}
Akmal, A., Pandharipande, V.~R., \& Ravenhall, D.~G. 1998, Phys. Rev. C, 58,
  1804

\bibitem[{Alsing {et~al.}(2018)Alsing, Silva, \& Berti}]{ns_masses}
Alsing, J., Silva, H.~O., \& Berti, E. 2018, MNRAS, 478, 1377

\bibitem[{{Antoniadis} {et~al.}(2013){Antoniadis}, {Freire}, {Wex}, {Tauris},
  {Lynch}, {van Kerkwijk}, {Kramer}, {Bassa}, {Dhillon}, {Driebe}, {Hessels},
  {Kaspi}, {Kondratiev}, {Langer}, {Marsh}, {McLaughlin}, {Pennucci}, {Ransom},
  {Stairs}, {van Leeuwen}, {Verbiest}, \& {Whelan}}]{Antoniadis2013}
{Antoniadis}, J., {Freire}, P. C.~C., {Wex}, N., {et~al.} 2013, Science, 340,
  448

\bibitem[{{Chetlur} {et~al.}(2014){Chetlur}, {Woolley}, {Vandermersch},
  {Cohen}, {Tran}, {Catanzaro}, \& {Shelhamer}}]{cuDNN}
{Chetlur}, S., {Woolley}, C., {Vandermersch}, P., {et~al.} 2014, arXiv
  e-prints, 1410.0759

\bibitem[{{Chodos} {et~al.}(1974){Chodos}, {Jaffe}, {Johnson}, {Thorn}, \&
  {Weisskopf}}]{Chodos1974}
{Chodos}, A., {Jaffe}, R.~L., {Johnson}, K., {Thorn}, C.~B., \& {Weisskopf},
  V.~F. 1974, \prd, 9, 3471

\bibitem[{Chollet {et~al.}(2015)}]{chollet2015keras}
Chollet, F. {et~al.} 2015, Keras, \url{https://keras.io}

\bibitem[{{Cromartie} {et~al.}(2020){Cromartie}, {Fonseca}, {Ransom},
  {Demorest}, {Arzoumanian}, {Blumer}, {Brook}, {DeCesar}, {Dolch}, {Ellis},
  {Ferdman}, {Ferrara}, {Garver-Daniels}, {Gentile}, {Jones}, {Lam}, {Lorimer},
  {Lynch}, {McLaughlin}, {Ng}, {Nice}, {Pennucci}, {Spiewak}, {Stairs},
  {Stovall}, {Swiggum}, \& {Zhu}}]{Cromantie2020}
{Cromartie}, H.~T., {Fonseca}, E., {Ransom}, S.~M., {et~al.} 2020, Nature
  Astronomy, 4, 72

\bibitem[{{De} {et~al.}(2018){De}, {Finstad}, {Lattimer}, {Brown}, {Berger}, \&
  {Biwer}}]{2018PhRvL.121i1102D}
{De}, S., {Finstad}, D., {Lattimer}, J.~M., {et~al.} 2018, \prl, 121, 091102

\bibitem[{Demorest {et~al.}(2010)Demorest, Pennucci, Ransom, Roberts, \&
  Hessels}]{Demorest2010}
Demorest, P., Pennucci, T., Ransom, S., Roberts, M., \& Hessels, J. 2010,
  Nature, 467, 1081

\bibitem[{{Douchin} \& {Haensel}(2001)}]{DouchinH2001}
{Douchin}, F. \& {Haensel}, P. 2001, \aap, 380, 151

\bibitem[{Fasano {et~al.}(2019)Fasano, Abdelsalhin, Maselli, \&
  Ferrari}]{Fasano_2019}
Fasano, M., Abdelsalhin, T., Maselli, A., \& Ferrari, V. 2019, Phys. Rev.
  Lett., 123, 141101

\bibitem[{{Ferreira} \& {Provid{\^e}ncia}(2019)}]{2019arXiv191005554F}
{Ferreira}, M. \& {Provid{\^e}ncia}, C. 2019, arXiv e-prints, arXiv:1910.05554

\bibitem[{Flanagan \& Hinderer(2008)}]{FlanaganH2008}
Flanagan, E.~E. \& Hinderer, T. 2008, Phys. Rev. D, 77, 021502

\bibitem[{{Fonseca} {et~al.}(2016){Fonseca}, {Pennucci}, {Ellis}, {Stairs},
  {Nice}, {Ransom}, {Demorest}, {Arzoumanian}, {Crowter}, {Dolch}, {Ferdman},
  {Gonzalez}, {Jones}, {Jones}, {Lam}, {Levin}, {McLaughlin}, {Stovall},
  {Swiggum}, \& {Zhu}}]{Fonseca2016}
{Fonseca}, E., {Pennucci}, T.~T., {Ellis}, J.~A., {et~al.} 2016, \apj, 832, 167

\bibitem[{{Fujimoto} {et~al.}(2018){Fujimoto}, {Fukushima}, \&
  {Murase}}]{2018PhRvD..98b3019F}
{Fujimoto}, Y., {Fukushima}, K., \& {Murase}, K. 2018, \prd, 98, 023019

\bibitem[{Fujimoto {et~al.}(2020)Fujimoto, Fukushima, \&
  Murase}]{PhysRevD.101.054016}
Fujimoto, Y., Fukushima, K., \& Murase, K. 2020, Phys. Rev. D, 101, 054016

\bibitem[{Goodfellow {et~al.}(2016)Goodfellow, Bengio, \&
  Courville}]{Goodfellow2016}
Goodfellow, I., Bengio, Y., \& Courville, A. 2016, Deep Learning (The MIT
  Press)

\bibitem[{Goriely {et~al.}(2010)Goriely, Chamel, \&
  Pearson}]{PhysRevC.82.035804}
Goriely, S., Chamel, N., \& Pearson, J.~M. 2010, Phys. Rev. C, 82, 035804

\bibitem[{{Haegel} \& {Husa}(2019)}]{2019arXiv191101496H}
{Haegel}, L. \& {Husa}, S. 2019, arXiv e-prints, arXiv:1911.01496

\bibitem[{{Haensel} \& {Pichon}(1994)}]{HaenselP1994}
{Haensel}, P. \& {Pichon}, B. 1994, \aap, 283, 313

\bibitem[{{Haensel} {et~al.}(2007){Haensel}, {Potekhin}, \&
  {Yakovlev}}]{HaenselPY2007}
{Haensel}, P., {Potekhin}, A.~Y., \& {Yakovlev}, D.~G. 2007, {Neutron Stars 1 :
  Equation of State and Structure}, Vol. 326 (New York, USA: Springer),
  pp.1--619

\bibitem[{Hernandez~Vivanco {et~al.}(2019)Hernandez~Vivanco, Smith, Thrane,
  Lasky, Talbot, \& Raymond}]{PhysRevD.100.103009}
Hernandez~Vivanco, F., Smith, R., Thrane, E., {et~al.} 2019, Phys. Rev. D, 100,
  103009

\bibitem[{Hinton \& Zemel(1993)}]{10.5555/2987189.2987190}
Hinton, G.~E. \& Zemel, R.~S. 1993, in Proceedings of the 6th International
  Conference on Neural Information Processing Systems, NIPS’93 (San
  Francisco, CA, USA: Morgan Kaufmann Publishers Inc.), 3–10

\bibitem[{Holt \& Lim(2019)}]{Holt_2019}
Holt, J.~W. \& Lim, Y. 2019, AIP Conference Proceedings, 2127, 020019

\bibitem[{{Kingma} \& {Ba}(2014)}]{ADAM2014}
{Kingma}, D.~P. \& {Ba}, J. 2014, arXiv e-prints, arXiv:1412.6980

\bibitem[{Kramer(1991)}]{Kramer1991NonlinearPC}
Kramer, M.~A. 1991, AIChE Journal, 37, 233

\bibitem[{{Love}(1911)}]{Love1911}
{Love}, A.~E.~H. 1911, {Some Problems of Geodynamics} (Univ. Press Cambridge)

\bibitem[{{Maggiore} {et~al.}(2020){Maggiore}, {Van Den Broeck}, {Bartolo},
  {Belgacem}, {Bertacca}, {Bizouard}, {Branchesi}, {Clesse}, {Foffa},
  {Garc{\'\i}a-Bellido}, {Grimm}, {Harms}, {Hinderer}, {Matarrese}, {Palomba},
  {Peloso}, {Ricciardone}, \& {Sakellariadou}}]{2020JCAP...03..050M}
{Maggiore}, M., {Van Den Broeck}, C., {Bartolo}, N., {et~al.} 2020, \jcap,
  2020, 050

\bibitem[{{Miller} {et~al.}(2019){Miller}, {Lamb}, {Dittmann}, {Bogdanov},
  {Arzoumanian}, {Gendreau}, {Guillot}, {Harding}, {Ho}, {Lattimer}, {Ludlam},
  {Mahmoodifar}, {Morsink}, {Ray}, {Strohmayer}, {Wood}, {Enoto}, {Foster},
  {Okajima}, {Prigozhin}, \& {Soong}}]{Miller2019}
{Miller}, M.~C., {Lamb}, F.~K., {Dittmann}, A.~J., {et~al.} 2019, \apjl, 887,
  L24

\bibitem[{Nickolls {et~al.}(2008)Nickolls, Buck, Garland, \& Skadron}]{cuda}
Nickolls, J., Buck, I., Garland, M., \& Skadron, K. 2008, Queue, 6, 40

\bibitem[{Oppenheimer \& Volkoff(1939)}]{OppenheimerV1939}
Oppenheimer, J.~R. \& Volkoff, G.~M. 1939, Phys. Rev., 55, 374

\bibitem[{{Press} {et~al.}(1992){Press}, {Teukolsky}, {Vetterling}, \&
  {Flannery}}]{Press1992}
{Press}, W.~H., {Teukolsky}, S.~A., {Vetterling}, W.~T., \& {Flannery}, B.~P.
  1992, {Numerical recipes in FORTRAN. The art of scientific computing}
  (Cambridge University Press; 2 edition)

\bibitem[{Raithel {et~al.}(2016)Raithel, Özel, \& Psaltis}]{Raithel_2016}
Raithel, C.~A., Özel, F., \& Psaltis, D. 2016, ApJ, 831, 44

\bibitem[{{Riley} {et~al.}(2019){Riley}, {Watts}, {Bogdanov}, {Ray}, {Ludlam},
  {Guillot}, {Arzoumanian}, {Baker}, {Bilous}, {Chakrabarty}, {Gendreau},
  {Harding}, {Ho}, {Lattimer}, {Morsink}, \& {Strohmayer}}]{Riley2019}
{Riley}, T.~E., {Watts}, A.~L., {Bogdanov}, S., {et~al.} 2019, \apjl, 887, L21

\bibitem[{Samuel(1959)}]{Samuel:1959}
Samuel, A.~L. 1959, IBM J. Res. Dev., 3, 210

\bibitem[{{Sieniawska} {et~al.}(2019){Sieniawska}, {Turcza{\'n}ski}, {Bejger},
  \& {Zdunik}}]{2019A&A...622A.174S}
{Sieniawska}, M., {Turcza{\'n}ski}, W., {Bejger}, M., \& {Zdunik}, J.~L. 2019,
  \aap, 622, A174

\bibitem[{Steiner {et~al.}(2010)Steiner, Lattimer, \& Brown}]{Steiner_2010}
Steiner, A.~W., Lattimer, J.~M., \& Brown, E.~F. 2010, ApJ, 722, 33

\bibitem[{Steiner {et~al.}(2013)Steiner, Lattimer, \& Brown}]{Steiner_2013}
Steiner, A.~W., Lattimer, J.~M., \& Brown, E.~F. 2013, ApJ, 765, L5

\bibitem[{Tolman(1939)}]{Tolman1939}
Tolman, R.~C. 1939, Phys. Rev., 55, 364

\bibitem[{{Tooper}(1965)}]{Tooper1965}
{Tooper}, R.~F. 1965, \apj, 142, 1541

\bibitem[{Traversi {et~al.}(2020)Traversi, Char, \&
  Pagliara}]{Traversi2020BayesianIO}
Traversi, S., Char, P., \& Pagliara, G. 2020, arXiv: High Energy Astrophysical
  Phenomena

\bibitem[{Van~Oeveren \& Friedman(2017)}]{VanOeverenF2017}
Van~Oeveren, E.~D. \& Friedman, J.~L. 2017, Phys. Rev. D, 95, 083014

\bibitem[{Wade {et~al.}(2014)Wade, Creighton, Ochsner, Lackey, Farr,
  Littenberg, \& Raymond}]{Wade:2014vqa}
Wade, L., Creighton, J. D.~E., Ochsner, E., {et~al.} 2014, Phys. Rev., D89,
  103012

\bibitem[{{Zdunik}(2000)}]{Zdunik2000}
{Zdunik}, J.~L. 2000, \aap, 359, 311

\bibitem[{{Zhang} {et~al.}(2019){Zhang}, {Qi}, \& {Wang}}]{2019arXiv190902274Z}
{Zhang}, N.-B., {Qi}, B., \& {Wang}, S.-Y. 2019, arXiv e-prints,
  arXiv:1909.02274

\end{thebibliography}

\end{document}